\documentclass[prb,twocolumn,superscriptaddress, floatfix,nobibnotes]{revtex4}
\usepackage{amsmath,,graphicx,epstopdf,color,bm,soul,upgreek, mathtools,siunitx}
\newcommand{\add}[1]{\textcolor{blue}{\uline{#1}}}
\begin{document}
\renewcommand{\add}[1]{\textcolor{blue}{{#1}}}
\newcommand{\delete}[1]{\textcolor{red}{#1}}

\title{Propagative oscillations in co-directional polariton waveguide couplers}

\author{J. Beierlein}
\email{johannes.beierlein@uni-wuerzburg.de}
\affiliation{Technische Physik, Wilhelm-Conrad-R\"ontgen-Research Center for Complex
Material Systems, and  W\"urzburg-Dresden Cluster of Excellence ct.qmat, Universit\"at W\"urzburg, Am Hubland, D-97074 W\"urzburg,
Germany}

\author{E. Rozas \footnote{J. Beierlein and E. Rozas contributed to this work equally.}}
\affiliation{Departamento de F\'{i}sica de Materiales, Instituto Nicol\'{a}s Cabrera, Universidad Aut\'{o}noma de Madrid,  28049 Madrid, Spain}

\author{O. A. Egorov} 
\affiliation{Institute of Condensed Matter Theory and Optics Friedrich-Schiller-Universit\"at Jena, Max-Wien-Platz 1, D-07743 Jena, Germany}

\author{M. Klaas}%
\affiliation{Technische Physik, Wilhelm-Conrad-R\"ontgen-Research Center for Complex
Material Systems, and  W\"urzburg-Dresden Cluster of Excellence ct.qmat, Universit\"at W\"urzburg, Am Hubland, D-97074 W\"urzburg,
Germany}

\author{A. Yulin}%
\affiliation{National Research University of Information Technologies, Mechanics and Optics (ITMO University), Saint-Petersburg 197101, Russia}

\author{H. Suchomel}%
\affiliation{Technische Physik, Wilhelm-Conrad-R\"ontgen-Research Center for Complex
Material Systems, and  W\"urzburg-Dresden Cluster of Excellence ct.qmat, Universit\"at W\"urzburg, Am Hubland, D-97074 W\"urzburg,
Germany}

\author{T.H. Harder}%
\affiliation{Technische Physik, Wilhelm-Conrad-R\"ontgen-Research Center for Complex
Material Systems, and  W\"urzburg-Dresden Cluster of Excellence ct.qmat, Universit\"at W\"urzburg, Am Hubland, D-97074 W\"urzburg,
Germany}

\author{M. Emmerling}%
\affiliation{Technische Physik, Wilhelm-Conrad-R\"ontgen-Research Center for Complex
Material Systems, and  W\"urzburg-Dresden Cluster of Excellence ct.qmat, Universit\"at W\"urzburg, Am Hubland, D-97074 W\"urzburg,
Germany}

\author{M. D. Mart\'{i}n}
\affiliation{Departamento de F\'{i}sica de Materiales, Instituto Nicol\'{a}s Cabrera, Universidad Aut\'{o}noma de Madrid, 28049 Madrid, Spain}

\author{I.A. Shelykh}%
\affiliation{Faculty of Physics and Engineering, ITMO University, 197101 St. Petersburg, Russia}
\affiliation{Science Institute, University of Iceland, IS-107 Reykjavik, Iceland}

\author{C. Schneider}%
\affiliation{Technische Physik, Wilhelm-Conrad-R\"ontgen-Research Center for Complex
Material Systems, and  W\"urzburg-Dresden Cluster of Excellence ct.qmat, Universit\"at W\"urzburg, Am Hubland, D-97074 W\"urzburg,
Germany}
\affiliation{Institute of Physics, University of Oldenburg, D-26129 Oldenburg, Germany}

\author{U. Peschel} 
\affiliation{Institute of Condensed Matter Theory and Optics Friedrich-Schiller-Universit\"at Jena, Max-Wien-Platz 1, D-07743 Jena, Germany}

\author{L. Vi\~{n}a}
\email{luis.vina@uam.es}
\affiliation{Departamento de F\'{i}sica de Materiales, Instituto Nicol\'{a}s Cabrera, Universidad Aut\'{o}noma de Madrid, 28049 Madrid, Spain}
\affiliation{Instituto de F\'{i}sica de la Materia Condensada, Universidad Aut\'{o}noma de Madrid, 28049 Madrid, Spain }

\author{S. H\"ofling}
\affiliation{Technische Physik, Wilhelm-Conrad-R\"ontgen-Research Center for Complex
Material Systems, and  W\"urzburg-Dresden Cluster of Excellence ct.qmat, Universit\"at W\"urzburg, Am Hubland, D-97074 W\"urzburg,
Germany}

\affiliation{SUPA, School of Physics and Astronomy, University of St Andrews, St Andrews
KY16 9SS, United Kingdom}

\author{S. Klembt}
\email{sebastian.klembt@physik.uni-wuerzburg.de}
\affiliation{Technische Physik, Wilhelm-Conrad-R\"ontgen-Research Center for Complex
Material Systems, and  W\"urzburg-Dresden Cluster of Excellence ct.qmat, Universit\"at W\"urzburg, Am Hubland, D-97074 W\"urzburg,
Germany}

\begin{abstract}
{We report on novel exciton-polariton routing devices created to study and purposely guide light-matter particles in their condensate phase. In a co-directional coupling device, two waveguides are connected by a partially etched section which facilitates tunable coupling of the adjacent channels. This evanescent coupling of the two macroscopic wavefunctions in each waveguide reveals itself in real space oscillations of the condensate. This Josephson-like oscillation has only been observed in coupled polariton traps so far. Here, we report on a similar coupling behavior in a controllable, propagative waveguide-based design. By \add{controlling the gap width, channel length or the propagation energy}, the exit port of the polariton flow can be chosen. This co-directional polariton device is a passive and scalable coupler element that can serve in compact, next generation logic architectures.}
\end{abstract}

\maketitle

\indent 
Photonic circuits rely on a variety of fiber-based optical elements for their functionality, which allow easy routing and filtering of the signals; the main drawback of purely photonic schemes for logic operations, however, is a lack of self interaction for very efficient switching \cite{Miller2010}. The remarkable advances in exciton-polariton physics are a result of the progressing control of high-quality microcavities, in which quantum well excitons and cavity photon modes couple strongly to form new hybrid light-matter eigenstates \cite{Weisbuch1992}. Polaritons exhibit a condensate regime at higher densities  \cite{Kasprzak2006} with emission properties similar to those of a traditional laser, without having to rely on population inversion \cite{Imamoglu1996}. This macroscopic quantum state, or quantum fluid of light \cite {Carusotto2013, Amo09}, can propagate over  macroscopic distances for high-quality samples \cite{Nelsen2013}. Furthermore, polaritons can be excited, confined  and therefore guided in waveguide structures \cite{Wertz2012}. Propitiously, the excitonic fraction of the polariton condensate is responsible for the observation of strong nonlinear interaction effects \cite{Vladimirova2010,Matutano2019,Delteil2019}, the photonic fraction allows for typical photonic benefits like a fast propagation velocity. Due to the combination of these two aspects, a variety of next generation devices based on polaritons can be envisioned \cite{Sanvitto2016}. Especially the possibility to use polaritons as information carriers in logic architectures has been addressed theoretically \cite{Liew2008} and experimentally, in proof of concept devices \cite{Amo2010,Ballarini2012,Gao2012,Nguyen2013,Suchomel2017,Anton2012,Anton2013,Winkler2017}. 
Quite recently, these ideas have been rekindled by room temperature experiments demonstrating coherent polariton propagation in perovskite waveguides \cite{Su2018} and a room-temperature organic polariton transistor \cite{Zasedatelev2019}.\\
Passive routing elements are essential in polariton logic architectures to make full use of the system as a low power consumption \cite{Deng2003} coherent light source. Basic routing effects have been achieved and predicted for polaritons  \cite{Flayac2013, Marsault2015}, which show some functionality but are mainly based on active optical control. To this end, we demonstrate a new polariton device in a  co-directional router, harnessing a Josephson-like oscillation effect in real space, which could feasibly be scaled and does not need active external control.\\ 
Josephson oscillations \cite{Josephson1962} occur when two quantum states are coupled by a transmissive barrier and were first demonstrated in superconductors \cite{Anderson1966}. Similar effects have been observed in atomic Bose Einstein condensates \cite{Levy2007, Smerzi1997, Gati2006, Gati2007, Albiez2005, Cataliotti2001} for which the interaction between the particles is crucial \cite{Legget2001} to observe different interaction dependent regimes of coupling. For exciton-polaritons this effect has first been observed in a naturally occurring disorder double potential well formed during sample growth \cite{Lagoudakis2010} and later in a dimer micropillar arrangement \cite{Abbarchi2013}, even achieving a regime where the interaction plays a crucial role in the time dynamics of these zero-dimensional systems \cite{Adiyatullin2017}.

\begin{figure}[htb!]
\centering
\includegraphics[width=0.98\linewidth]{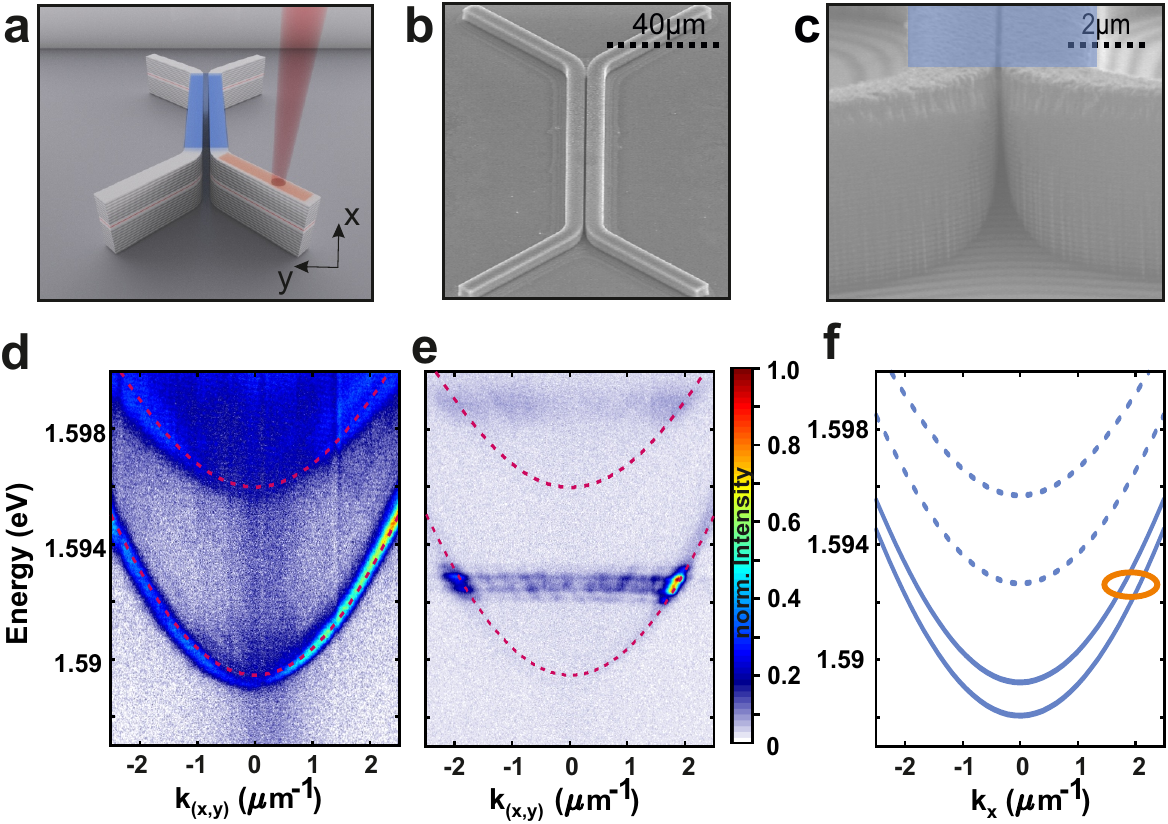} 
\caption{(a) Device schematic with indicated laser excitation (red), incoupler region(orange) and the coupling region (blue) along the $x$ axis. (b) Top view SEM image of a co-directional polariton coupler. (c) Zoom-in highlighting the coupling region(blue) and the gap between the two waveguides. Here, the cavity and a varying number of mirror pairs are still intact. (d) Measured waveguide dispersion below and (e) above polariton condensation threshold at the input port region (orange) \add{parallel to the waveguide}, \add{which is angled $45^\circ$ to the $x$ axis. (f) Calculated dispersion below condensation threshold at the \add{coupling region (blue)} along the $x$ axis.}}
\label{fig1}
\end{figure}

Our new coupler device consists of extended one-dimensional channels which allow observation of the oscillations in the spatial domain. 
We use a new, specifically tailored fabrication technique where the top mirrors between the waveguides are partially etched to realise controlled coupling and thus oscillatory exchange of the polariton population. This technique allows routing of polaritons depending on controllable device parameters and the resulting tunable coupling between the waveguides.

\indent 
In this work, dedicated to a proof of concept, we use high-quality GaAs-based microcavities, benefiting from mature fabrication techniques \cite{Schneider2016}. For the co-directional couplers with waveguide coupling achieved by partially etched mirrors we use molecular beam epitaxially grown microcavities featuring 27 distributed Bragg reflector (DBR) pairs of AlAs/Al\textsubscript{0.2}Ga\textsubscript{0.8}As in the bottom and 23 in the top DBR. Three stacks of four GaAs quantum wells with a width of 7\,nm are placed in the AlAs $ \lambda/2$ cavity at the anti-nodes of the electric field. The vacuum Rabi splitting determined by white light reflection measurements is 13.9\,meV. The quality factor $ Q \approx 5000$ was determined at low power excitation.

Sample processing was done via a specially developed reactive ion etching (RIE) process. The first step consists of an electron beam exposure of a polymethyl methacrylate photoresist and subsequent development. Later, a metal layer of calibrated thickness is evaporated on the sampled followed by a lift-off process. After the lift-off process the sample is etched. Due to protection by the predefined metal layers, the sample is only etched at the exposed positions, which allows the fabrication of the waveguide structures. Due to the proximity of the structures and the anisotropic etching behaviour of RIE, the etching rate between the waveguides is slower, leaving a certain number of mirror pairs untouched and resulting in a rising flank at the etch edges. These left-over mirror pairs between the waveguides facilitate evanescent, photonic coupling. The area around the defined waveguide structures is nearly etched through the bottom DBR and therefore facilitates strong photonic confinement.

Fig.~\ref{fig1}(a) shows a sketch of the intended structure, highlighting an excitation scheme (red), the incoupler region (orange) and the propagation direction $x$ \add{along} the coupling area (blue) . A scanning electron microscopy image of the full device as well as a zoomed view of the coupling region, stressing the fabrication-induced narrow gap between the structures, are presented in Figs. 1(b) and (c), respectively. The incoupler, which is angled 45$^{\circ}$  to the propagation direction $x$ (see Fig.~\ref{fig1}(a)), has a width of $2\,\mu$m and a length of $40 \,\mu$m. Figs.~\ref{fig1} (d) and (e) depict the energy  dispersion of the ground state along the incoupler region in the linear regime and above threshold, respectively.\add{ The calculated dispersion along the coupler is shown in Fig.~\ref{fig1}(f). }  \\
The experiments have been carried out with two photoluminescence setups, the first capable of both Fourier- and real space emission detection while the second one was used for streak camera measurements. Excitation was provided using a tuneable Ti:Sa-Lasers with 10\,ps and 2\,ps pulse lengths set at the wavelength of a high energetic Bragg mode of the microcavity at each structure for efficient injection. The excitation was mechanically chopped with a ratio of 1:12 to prevent sample heating. Additionally, a tomography technique using motorized lenses was implemented to allow for energy selective imaging. The pump spot was focused via a microscope objective with NA = 0.42 to a diameter of $\sim 3$ $\upmu$m.

\begin{figure} 
\centering
\includegraphics[width=0.9\linewidth]{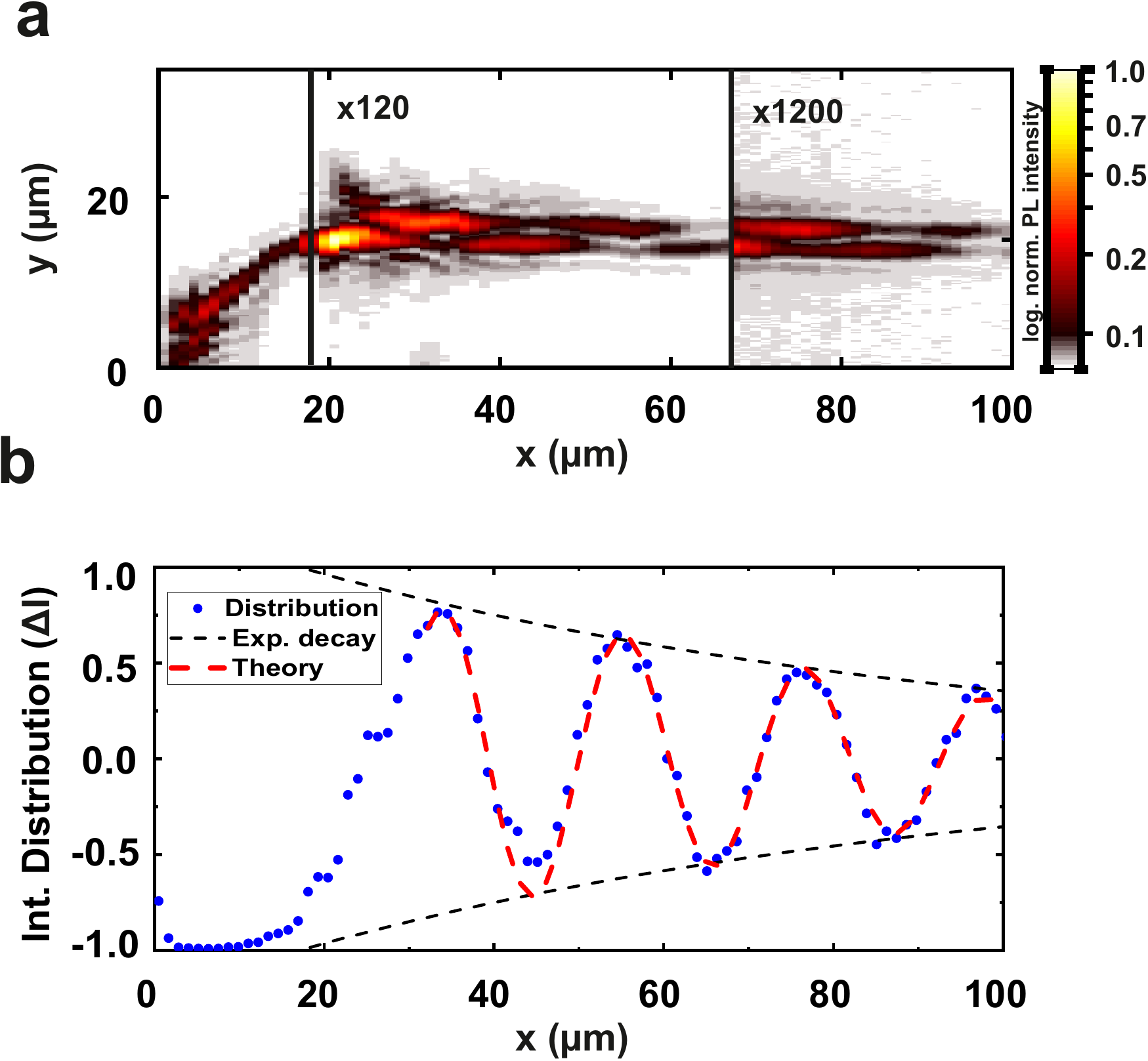} 
\caption{(a) Energy-resolved real space photoluminescence signal for a coupler device with a gap size of 200\,nm and a coupler region length of 100\,$\mu$m at \add{ $E = 1.5924-1.5930$\,eV}. Polaritons are injected non-resonantly in the lower left incoupler and exhibit a distinct oscillatory behavior between the two waveguides in the coupler region (right of the black line). (b) Polariton photoluminescence distribution  \add{ $\Delta I = (I_{ \text{top}}-I_{ \text{bot}}) / I_{ \text{tot}}$ } plotted for the lower (negative) and the upper waveguide (positive) showing the oscillation as well as the damping that is expected for a dissipative system. The red dashed line represents the results of theoretical fitting by Eq.~(\ref{Intensity}).}
\label{fig3}
\end{figure}

As an example, we show the characterization of a single waveguide with an excitation spot located at the center of the incoupler. The dispersion depicted in Fig.~\ref{fig1}(d) was measured parallel to the incoupler at a low excitation density of 0.1\,mW and features photoluminescence from two modes, characteristic for the waveguide's one dimensional confinement. These modes have been fitted with an approach from Ref. \cite{Tartakovskii.1998}.
This allows to extract a detuning of  \add{-20\,meV} with an exciton energy of 1.609\,eV, corresponding to photonic and excitonic fractions of\add{  90.6\% and 9.4\%}, respectively. 


To move on, an investigation of the non-linear regime in this structure, is represented in Fig.~\ref{fig1}(e), showing polaritons above threshold being expelled from the small laser pump spot \cite{Wertz2010} at a wave vector k\textsubscript{x} $\approx$ 1.9 $ \mu \text{m}^{-1}$. A detailed analysis of the emission intensity and energy as a function of the excitation power revealing a lasing threshold behavior and a continuous blueshift of 4\,meV  is presented in the supplementary materials, confirming polariton interaction effects\cite{Ciuti.1998}.



Let us now report on the main subject of the work, namely the oscillations in co-directional couplers. To this end, Fig.~\ref{fig3} (a) depicts a logarithmic color-coded real space image of the energy-resolved photoluminescence from two adjacent waveguides at an injection power of 1\,mW, corresponding to the propagating condensate regime for\add{ $E = 1.5924-1.5930$\,eV}. The gap between the two waveguides is 200\,nm wide and the coupling area is $100\,\mu$m long.  The emission from the region of the excitation spot (on the bottom left incoupler) is attenuated by a neutral density filter,\add{ while the far propagating condensate is amplified to compensate for its disspative nature.} A clear oscillation pattern is observed upon propagation of the condensate into the coupling region of the two coupled waveguides. In panel (b), a quantitative analysis of the oscillation by plotting the  intensity distribution $\Delta I = (I_{ \text{top}}-I_{ \text{bot}}) / I_{ \text{tot}}  $  between the two waveguides, where $I_{ \text{top}},I_{ \text{bot}}$ and $ I_{ \text{tot}}$ denote the intensities in the top, bottom and both waveguides is presented. Therefore, $\Delta I = 0$ corresponds to an equal distribution of the population. Due to continuous polariton decay, we observe an exponential spatial decay of the distribution characterized by a decay constant of $\tau_x = 80\,\mu \text{m}^{-1}$. From this distribution a spatial oscillation period of $\sim 20\,\mu \text{m}$ can be extracted.


\begin{figure}
\centering
\includegraphics[width=1\linewidth]{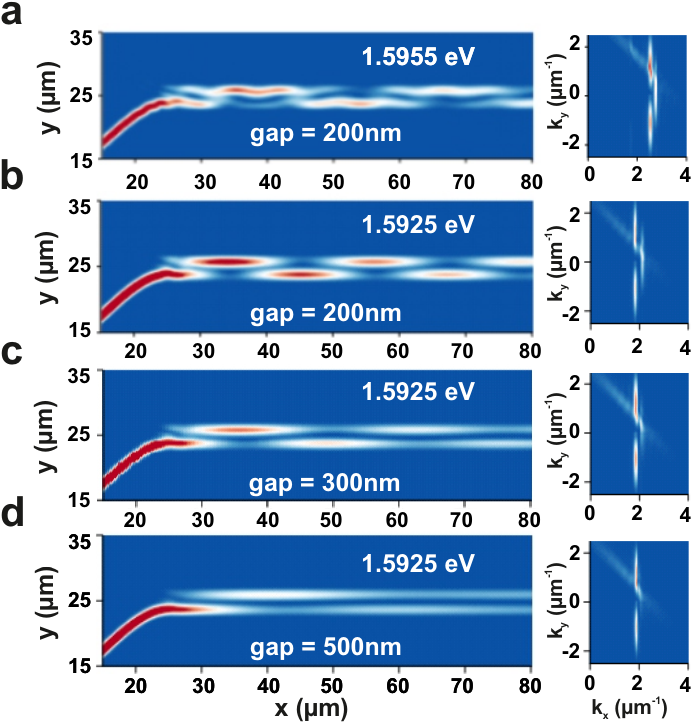}
\caption{Propagation dynamics of polaritons calculated within the model Eqs.~\ref{eq:phot} \& \ref{eq:exc} for coupled waveguides with separation gap sizes of (a,b) 200\,nm, (c) 300\,nm, and (d) 500\,nm. \add{While the dynamics for (a) is at an energy of 1.5955\,eV (b-d) are at 1.5925\,eV. For small gap sizes (a-c) a clear oscillatory behaviour is observed. In addition, a change in the oscillation period is observable for two different energies and the same gap size. } The insets on the right hand sides show the respective distribution in momentum space (on the parameter plane $k_y$ and $k_x$). Antisymmetric and symmetric modes of the coupler are visible. }
\label{fig4}
\end{figure}

The governing physics behind the observed oscillation dynamics can be understood within a slowly- varying amplitude approach for the modes of the coupled waveguides. Since the excitation occurs above condensation threshold,  the ballistic polaritons  have a well-defined frequency. Thus we focus on spatial dynamics along the propagation axis $x$ considering a monochromatic case.          
In the low-intensity limit the dynamics can be modelled via two coupled equations for amplitudes $A_{1,2}$ in the coupled waveguides 
\begin{equation}
i \frac{\partial}{\partial x} A_1(x)= - \kappa A_1(x)- J A_2(x),
\label{eq1}
\end{equation}
\begin{equation}
i \frac{\partial}{\partial x} A_2(x)= - \kappa A_2(x)- J  A_1(x).
\label{eq2}
\end{equation}
where $J \equiv J^{\prime} +i J^{\prime{\prime}} $ is the complex coupling constant which is governed by the width and depth of the gap between the coupled waveguides. The complex wavevector $\kappa=k+i \alpha$ characterizes the propagation and damping of guided modes in separated waveguides (in the following $\alpha > J^{\prime{\prime}} >0)$. 


In order to find the super-modes of the coupler we look for a solution of Eqs.~(\ref{eq1}) and (\ref{eq2}) in the form $A_{1,2}(x)=a_{1,2}\exp(i \beta x)$. By diagonalizing the system one finds two eigenmodes, namely symmetric $a_{1}=a_{2}$ and antisymmetric $a_{1}=-a_{2}$ one, with the propagation constants  $\beta_+=\kappa + J^{\prime} +i J^{\prime{\prime}} $ and $\beta_-=\kappa - J^{\prime} -i J^{\prime{\prime}} $, respectively. It is worth mentioning that, due to the imaginary parts of the coupling constants, the \add{damping rates}  $Im (\beta_{\pm})=\alpha \pm J^{\prime{\prime}}$ of these two modes \add{are} different. More precisely, the antisymmetric mode has smaller propagation losses.

If only one waveguide is excited (with an amplitude $A$), the analytical solution for the mode dynamics can be easily   found as 
\begin{align}
\begin{split}
\label{Solution1}
A_1 &(x) = A e^{ik x-\alpha x}  \\ 
& \times \left( \cos (J^{\prime}x) \cosh (J^{\prime{\prime}}x) -i \sin (J^{\prime}x) \sinh (J^{\prime{\prime}}x) \right),
\end{split} \\
\label{Solution2}
\begin{split}
A_2 &(x)  =  A e^{ik x-\alpha x}   \\
& \times \left( i \sin (J^{\prime}x) \cosh (J^{\prime{\prime}}x)  - \cos (J^{\prime}x) \sinh (J^{\prime{\prime}}x) \right).
\end{split}
\end{align}

\add{ From the analytical solution~(\ref{Solution1},\ref{Solution2}) it is easy to derive a simple expression for the intensity distribution in the coupler }
\begin{align}
\label{Intensity}
\add{ \Delta I = (I_{ \text{1}}-I_{ \text{2}})/I_{ \text{tot}} = \cos(2J^{\prime}x)/ \cosh(2J^{\prime\prime}x).}
\end{align}

\add{This solution has a form of damped oscillations where polaritons transfer between the two channels with a spatial period $\pi/J^{\prime} =  21.36\,\mu m$ and a damping coefficient  given by the imaginary part of the coupling constant $2J^{\prime\prime}=0.021\,\mu m^{-1}$ }   [see Fig.~\ref{fig3} (b)] . The physical origin of the oscillations is the beating of two (symmetric and antisymmetric) eigenmodes. 

It is worth mentioning, that, due to the etching of the Bragg mirror between two channels, the effect of local losses becomes comparable with the polariton tunnelling dynamics and thus the imaginary part of the coupling cannot be neglected. For a non-zero imaginary part $J^{\prime{\prime}} > 0$, the symmetric mode decays faster and at propagation distances of the order of $ 1/J^{\prime{\prime}}$ it becomes much less intensive than the antisymmetric mode. This suppresses the mode beating at large propagation distances.

To underpin this rather qualitative analysis with a more thorough theoretical study, we performed numerical calculations in the frame of the mean-field model for 2D intracavity photons coupled strongly to the quantum well excitons \cite{Amo09, Carusotto2013}. This is a widely accepted approach for exciton-polariton dynamics in microcavities where the required waveguiding geometry can be accounted by inducing an appropriate potential for photons.  Neglecting polarization effects one obtains two coupled Schr\"odinger equations for the photonic field $\Psi_\text{C}$ and coherent excitons  $\Psi_\text{E}$ given as
\begin{equation}\begin{split}
\label{eq:phot}
\partial _t {\Psi_\text{C} } -  & \frac{i\hbar}{2m_{\text{c}}}  \nabla_{x,y} ^2{\Psi_\text{C} } + i V(x,y) {\Psi_\text{C} } \\& + \left[ {\add{\gamma_\text{c}}  - i\left( {{\omega _\text{p}} \add{- \delta} } \right)} \right]{\Psi_\text{C} }    = i{\Omega _\text{R}{\Psi_\text{E} } + \Psi_\text{C} (x,y) e^{ik_\text{p}x}},
\end{split}
\end{equation}
\begin{equation}\begin{split}
\label{eq:exc}
\partial _t{\Psi_\text{E} }  - \frac{i\hbar}{2m_{\text{E}}}  \nabla _{x,y} ^2{\Psi_\text{E} } + \left[ {\add{\gamma_\text{e}}  - i{\omega _\text{p}}} \right]{\Psi_\text{E}} = i{\Omega _\text{R}}{\Psi_\text{C}  \,  .}
\end{split}
\end{equation}

The complex amplitudes are obtained through a standard averaging procedure of the related creation or annihilation operators.\add{ $\gamma_c$ and $\gamma_e$  denote the cavity photon damping and dephasing rate of excitons, respectively. We note that, for cryogenic sample temperatures, the excitonic dephasing time can be comparable with the life-time of intracavity photons. Thus, without loss of generality, they are assumed to be equal $\hbar\gamma_c=\hbar\gamma_e=0.01\, $meV.} The effective photon mass in the planar region is given by $m_\text{c}=36.13 \times 10^{-6} m_\text{e}$ where $m_\text{e}$ is the free electron mass. The effective mass of excitons is $m_\text{E}\approx 10^5 m_\text{C}$.    $\Omega_{\text{R}}$ is the Rabi frequency which defines the Rabi splitting $2\hbar \Omega_{\text{R}}=13.9\,\text{meV}$. The photon - exciton detuning is given by the parameter \add{ $\hbar \delta = \hbar \omega_{c} - \hbar \omega_{e}=-20\,\text{meV}$  } where  $\omega_{\text{c}}$ is the cavity resonance frequency and \add{ $\omega_{e}$ } is the excitonic resonance.

The external photonic potential \add{ $V(x,y)=V_{\text{real}}(x,y)+i V_{\text{imag}}(x,y)$} defines the waveguide geometry induced by etching of the Bragg mirror. \add{ An appropriate profile of the potential was chosen  after a thorough fitting of the dispersion relation of the guided modes known from the experiment [see Figs.~\ref{fig1}(d) and (e)]. The real part of the best-fitted profile was found to be a super-Gauss $V_{\text{real}}(y)=V_{\text{re}}\cdot(1-\text{exp}(-y^8/s^8))$ with the depth $\hbar V_{\text{re}}= 40\, \text{meV}$ and the width of $s=2\mu m$. We also estimated the potential depth in the gap between two waveguides in the coupler region to be $6\, \text{meV}$ [see the calculated dispersion relation in Fig.~\ref{fig1}(f)].
The imaginary part of the  potential accounts for additional damping induced by etching of the Bragg mirror and was approximated by the expression  $V_{\text{imag}}(y)=-V_{\text{im}}\cdot(1-\text{exp}(-y^{16}/s_{\text{im}}^{16}))$ with $\hbar V_{\text{im}}= 5\, \text{meV}$ and $s_{\text{im}}=2.24\mu m$.}
 
Since we are interested in the propagation dynamics of polaritons with a well-defined frequency it is sufficient to consider  the case of coherent excitation by a pump beam \add{ $W_\text{p} (x,y)$} with a frequency $\omega$ and with a momentum $k_\text{p}$ such that the parameter $\omega_{\text{p}} = \omega - \omega_{0}$ describes the detuning of the pump frequency $\omega$ from the excitonic resonance.


\add{Figs.~\ref{fig4}(a) and (b) show examples of oscillation dynamics for two slightly different frequencies of the wave-packets. These oscillations are governed by interference of the symmetric and the antisymmetric modes of the coupler, which is clearly visible in the two-dimensional spectrum, shown in the insets to Figs.~\ref{fig4}. The insets in each panel represent the momentum space distribution of the propagating polaritons.  The substantial difference in the period of spatial oscillations can be explained by a non-equidistant momentum splitting between the mentioned above symmetric and antisymmetric modes within the dispersion relation of the coupler [see Fig.~\ref{fig1}(f)].  }

\add{
Further examples in Figs.~\ref{fig4}(c)-(d) show propagation dynamics in which the coupling strength is continuously weakened due to decreased wavefunction overlap via increased gap width. 
A clear change in the oscillation pattern is observed.
}

Note that, due to pronounced dissipative effects within the gap between waveguides, the antisymmetric mode, which has the lowest overlap with the gap region, dominates the spectrum for larger gaps and, as a result, oscillations disappear [see Fig.~\ref{fig4}(d)].    
\begin{figure}  [tb!]
\centering
\includegraphics[width=0.9\linewidth]{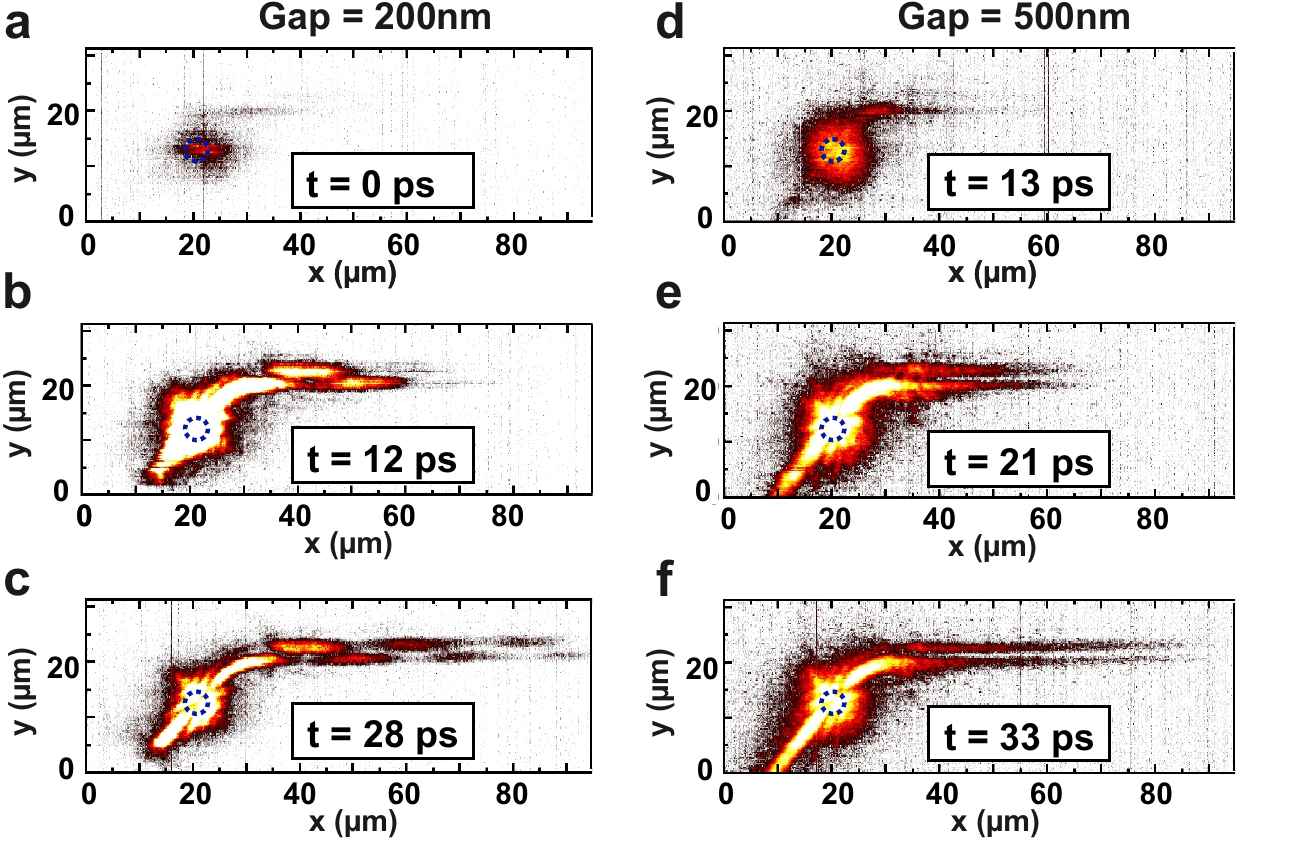} 
\caption{a-c) Time-resolved propagation of the polariton propagation in co-directional coupler devices with a gap size of (a-c) 200\,nm and (d-f) 500\,nm for 3 different times  during the propagation. While (a-c) show a clean oscillation between the two waveguides, no oscillation is visible in (d-f). \add{The pump area is indicated by a blue dashed circle.} }
\label{fig5}
\end{figure}

\begin{figure} [tb!]
\centering
\includegraphics[width=\linewidth]{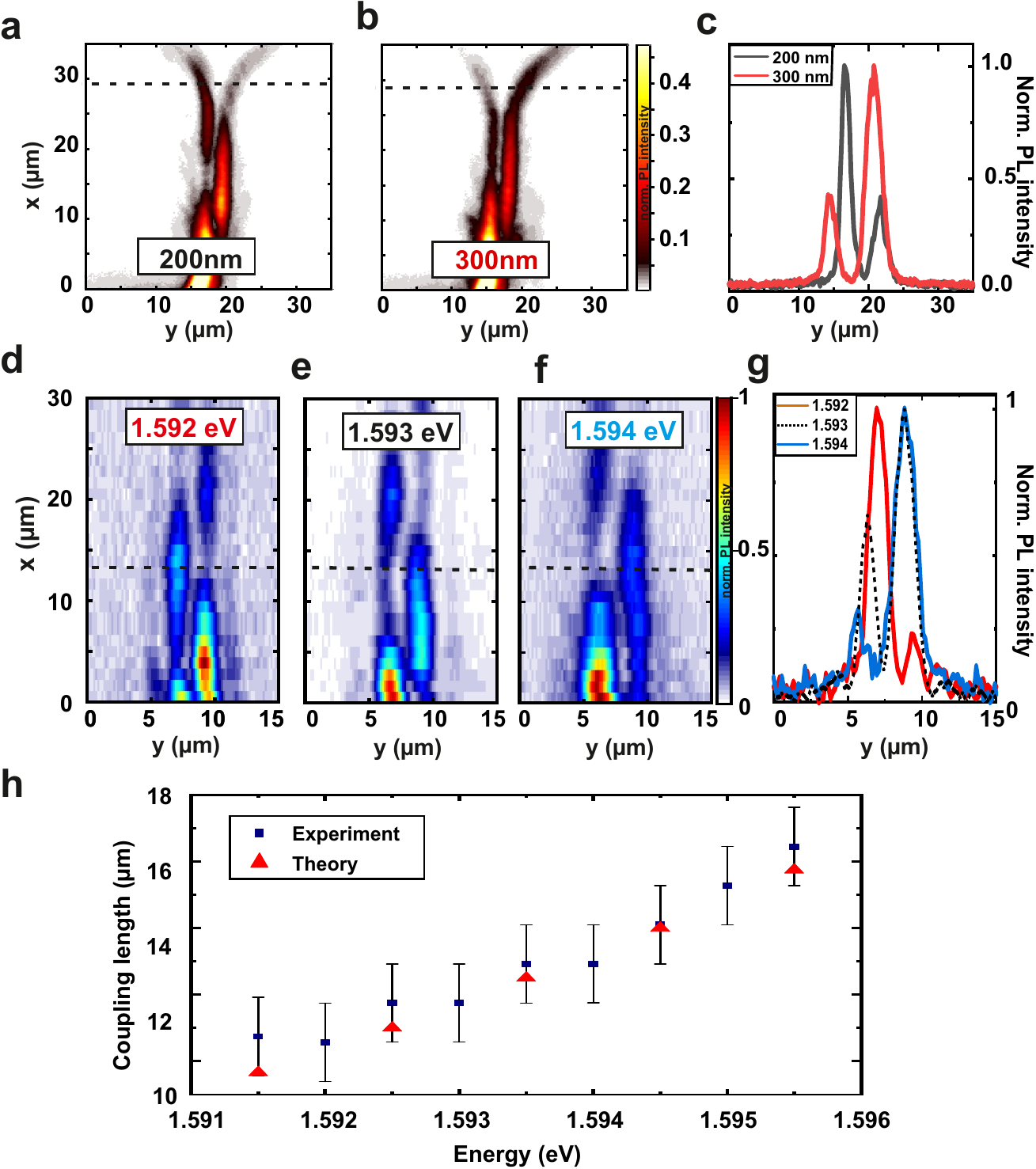} 
\caption{\add{Routing of a polariton condensate via a different coupling length (a-c) and different propagation energy (d-g). (Note  that the plot orientation has been turned by $90^\circ$ here for visual clarity). A coupler with a length of the coupling region of of 20\,$\mu$m and a gap size of (a) 200\,nm and (b) 300\,nm is shown. By reducing the coupling, the oscillation \add{period} and thus the dominant output port is changed. (c) Line profiles of the normalized intensities at $x=28\,\mu$m for (a) and (b). Here a routing ratio of the integrated intensities of 1.9 and 3.1 could be extracted, respectively. PL Emission for different energies (d-f) in a fixed region of the coupler is shown. A clear change in the oscillation \add{period} is observed. (g)  Normalized PL intensity line profiles of the emission at $x=13\,\mu$m. (h) Experimentally extracted coupling lengths as a function of the propagation energy and numerically calculated values.}}
\label{fig6}
\end{figure}

Now, in order to demonstrate the polariton dynamics in the co-directional coupler devices experimentally, we have performed energy- and time-resolved streak camera measurements using two devices with gap sizes of 200\,nm and 500\,nm. Using a streak camera with an overall time resolution of 10 ps, the PL has been measured up to 150 ps after the laser beam excites the structure at t=0. However, due to the fast dynamics of polaritons in this sample, the polariton propagation is only shown up to ∼30 ps.  The respective intensity patterns are plotted in Figs.~\ref{fig5} (a)-(c) and (d)-(f).\\
In Fig.~\ref{fig5} (a) at t=0\,ps, we observe the laser excitation spot on the lower left input coupler  from where polaritons are repulsively expelled into the coupling region. At t=12\,ps, the polaritons have finished the first full oscillation in the upper waveguide and are back in the lower one. Finally, after approximately 30\,ps, the polariton population has dissipated after a propagation length of ~100\,$\mu$m, underlining the excellent quality of the patterned microcavity structure. The oscillation \add{period} is $\sim 20\,\mu \text{m}$ in excellent agreement with the theoretical model. Figs.~\ref{fig5} (d)-(f) shows the temporal evolution in a system with a much larger gap of 500\,nm.   In this case, while there is some evanescent coupling to the upper waveguide, no pronounced oscillatory behavior is observed, again in excellent agreement with the theory presented in Fig.~\ref{fig4}.\\
From the experiments as well as from the theoretical model, we can infer that a variation in the coupling strength ultimately allows for a change of the oscillation \add{period}. This can now be used to choose the output-port of the polariton flow due to a specifically tailored channel length and etch depth. Fig.~\ref{fig6} shows the results of the experiments. We have used similar couplers with 200\,nm and 300\,nm gap size respectively, but with a reduced coupler region length of 20\,$\mu$m. While (a) shows the polariton flow leaving predominantly through the \add{left} outcoupler, the larger gap (weaker coupling and larger oscillation length) in (b) shows predominant coupling to the \add{right} arm. \add{In (c) the normalized line-profiles are plotted for Figs.~\ref{fig6}(a,b) at $x=28\,\mu$m. A clear change in output is visible and can be characterized by the ratio of the integrated routed intensities. These were extracted as 1.9 and 3.1 for Figs.~\ref{fig6}(a,b), respectively.
\add{Another way of changing the output port is by changing the energy of the polariton condensate. The emissions of a condensate for three different energies ranging from 1.592-1.594\,eV are shown in Figs.~\ref{fig6}(d-f). Changing the condensate propagation energy by 2\,meV, a change of the intensity from left to right can be observed. This shift is underlined by the line profiles in Fig.~\ref{fig6}(g). The detailed change of coupling lengths and the numerically calculated values for Figs.~\ref{fig6}(d-f) are depicted in the Fig.~\ref{fig6}(h).
}}

 Therefore we have shown that this device configuration allows co-directional routing to a predetermined exit-port via a Josephson-like oscillation effect in real space by engineering the \add{period} and coupling length.\\
In conclusion, we have demonstrated the possibility for passive polariton routing, which is easily scalable and integratable to large polariton based logic networks. We evidenced this by a precise control of the lithographically engineered photonic landscape, which allows for the observation of these oscillations in real space between polaritonic waveguides. Such detailed tailoring of the flow of quantum fluids of light paves the way to harness their non-linearity in next generation photonics. Furthermore, the basic understanding of the coupling of polariton waveguides \cite{Klaas2019,Rozas2020} is the necessary foundation for larger coupled waveguide arrays, comparable to those that have been implemented in Si-photonics for the demonstration of transport in a topological defect waveguide \cite{Redondo2016} or Floquet waveguides  for the study of topologically protected bound states in photonic parity-time symmetric crystals\cite{Weimann2017}. In this respect, our work opens a new route, to use polariton waveguides for polariton logic as well as for topological devices involving nonlinearity, gain, interactions and coherence, inherent to the polariton system.
\begin{acknowledgments}

The W\"urzburg and Jena group acknowledge financial support within the DFG projects SCHN1376/3-1, PE 523/18-1 and KL3124/2-1 . The W\"urzburg group is grateful for support from the state of Bavaria and within the  W\"urzburg-Dresden Cluster of Excellence ct.qmat. S.H. also
acknowledges support by the EPSRC “Hybrid Polaritonics” grant (EP/
M025330/1). The W\"urzburg group wants to thank Hugo Flayac for inspiring discussions in the early stage of this work.\\
The Madrid group acknowledges financial support of the Spanish MINECO Grant MAT2017-83722-R. E.R. acknowledges financial support from a Spanish FPI Scholarship No. BES-2015-074708.
I.A.Shelykh acknowledges financial support of Icelandic Science Foundation (project “Hybrid polaritonics”).  I.A.Shelykh and A.Yulin acknowledge Ministry of Science and Higher Education of the Russian Federation (Megagrant number 14.Y26.31.0015).

\end{acknowledgments}

\end{document}